\documentclass[useAMS,usenatbib]{mn2e}
\usepackage{aas_macros}
\usepackage[dvips]{graphicx}
\usepackage{amsmath,amssymb}
\title[The biasing of baryons on the cluster mass function and cosmological parameter estimation]{
The biasing of baryons on the cluster mass function and cosmological parameter estimation
\
}
\author[D. Martizzi et al.]{\parbox[t]{\textwidth}{Davide Martizzi$^{1,2}$\thanks{E-mail: martdav@physik.uzh.ch, davide.martizzi@berkeley.edu}, 
Irshad Mohammed$^{1}$, Romain Teyssier$^{1}$, Ben Moore$^{1}$}\vspace*{6pt}\\
$^{1}$Institute for Theoretical Physics, University of Zurich, CH-8057 Z\"urich, Switzerland\\
$^{2}$Department of Astronomy, University of California, Berkeley, CA 94720-3411, USA \\
}

\begin{document}

\maketitle

\label{firstpage}

\begin{abstract}

We study the effect of baryonic processes on the halo mass function in the galaxy cluster mass range using a catalogue of 153 high resolution cosmological hydrodynamical simulations performed with the AMR code {\scshape ramses}. We use the results of our simulations within a simple analytical model to gauge the effects of baryon physics on the halo mass function. Neglect of AGN feedback leads to a significant boost in the cluster mass function similar to that reported by other authors. However, including AGN feedback not only gives rise to systems that are similar to observed galaxy clusters, but they also reverse the global baryonic effects on the clusters. The resulting mass function is closer to the unmodified dark matter halo mass function but still contains a mass dependent bias at the 5--10\% level. These effects bias measurements of the cosmological parameters, such as $\sigma_{\rm 8}$ and $\Omega_{\rm m}$. For current cluster surveys baryonic effects are within the noise for current survey volumes, but forthcoming and planned large SZ, X-ray and multi-wavelength surveys will be biased at the percent level by these processes. The predictions for the halo mass function including baryonic effects need to be carefully studied with larger and improved simulations. However, simulations of full cosmological boxes with the resolution we achieve and including AGN feedback are still computationally challenging.
\end{abstract}


\section{Introduction}

In the hierarchical structure formation scenario, the abundance of dark matter halos within a representative volume of the universe is characterized by the halo mass function. Since the statistical properties of the cosmic density field are related to the underlying cosmological model, the halo mass function contains great deal of information about the cosmological parameters of our Universe. In particular, within the standard $\Lambda$ cold dark matter cosmological scenario, dark matter halos form from initial density peaks via gravitational instability. The resulting halo mass function is directly related to the statistical properties of the primordial density field. Early studies suggested that the halo mass function can be expressed as a universal function of $\sigma(M)$, the rms value of the density perturbations at a particular mass scale $M$ \citep{1974ApJ...187..425P, 1988MNRAS.235..715E, 1999MNRAS.308..119S, 2001MNRAS.323....1S}.

Cosmological N-body simulations have been extensively used to calibrate the halo mass function and to test whether this quantity is universal and how accurately it can be determined \citep{1997A&A...325..439A, 2001MNRAS.323....1S, 2001MNRAS.321..372J, 2003MNRAS.346..565R, 2005MNRAS.364.1105S, 2006ApJ...646..881W, 2007MNRAS.374....2R, 2008ApJ...688..709T, 2010MNRAS.403.1353C, 2013MNRAS.431.1866R}. These studies demonstrated that the mass function measured in cosmological N-body simulations deviates from universality at the 10\% level. Despite the importance of precise calibrations of the halo mass function and the ability to make accurate theoretical predictions, there is another issue that needs to be considered when studying the halo mass function: the effect of baryons on halo masses. This issue might be extremely relevant for the comparison of observationally determined halo mass functions and the prediction of cosmological simulations. In fact, present and next generation surveys such as Euclid, the Dark Energy Survey or eROSITA, are expected to measure halo masses and the halo mass function with an unprecedented percent level precision.

The effect of baryonic processes on the power spectrum and on the weak gravitational lensing shear signal has been studied in detail \citep{2004APh....22..211W, 2004ApJ...616L..75Z, 2006ApJ...640L.119J, 2008ApJ...672...19R, 2010MNRAS.405..525G, 2011MNRAS.415.3649V, 2011MNRAS.417.2020S, 2013arXiv1306.4686R}. A series of recent papers \citep{2009MNRAS.394L..11S, 2012MNRAS.423.2279C, 2013JCAP...04..022B} attempts to assess the importance of baryonic processes in shaping the properties of the halo mass function. The simulations performed by \cite{2009MNRAS.394L..11S} and \cite{2012MNRAS.423.2279C} show a significant increase of halo counts in the galaxy cluster mass range in the cases in which cooling and star formation are considered. This is caused by the condensation of baryons at the centres of dark matter halos which leads to the contraction of the matter distribution and an enhancement of their masses. However, none of the simulations used by these authors include AGN feedback, which is expected to be an important process in shaping the properties of galaxy clusters, especially of their mass distribution \citep{2010MNRAS.405.2161D, 2011MNRAS.414..195T, 2012MNRAS.422.3081M}. In this paper, we carry out a large suite of simulations that follow the formation of galaxy clusters over a range of masses. These are fully cosmological hydrodynamical simulations performed with the {\scshape ramses} code. The aim of this study is to explicitly study the impact of AGN feedback on halo masses and the effects of baryonic processes on the halo mass function at redshift $z=0$ and the subsequent biases this introduces in recovering the underlying cosmological parameters. 

The paper is organized as follows. In Section 2, we describe the numerical simulations. In Section 3, we describe the analytical formalism we adopt to compute the effect of baryons on the halo mass function. In Section 4, we discuss the measurements we perform on the simulations that will be used as an input for the analytical formalism. In Section 5, we show our predictions on the halo mass function. In Section 6, we test the impact of our predictions on the estimate of cosmological parameters. In Section 7, we summarize and discuss our results.

\section{The simulations}
\label{sec:num_methods}

We performed a set of 153 cosmological resimulations with the {\scshape ramses} code \citep{Teyssier:2002p451}. We assume the standard $\Lambda$CDM cosmological scenario with matter density parameter $\Omega_{\rm m}=0.272$, cosmological constant density parameter $\Omega_\Lambda=0.728$, baryonic matter density parameter $\Omega_{\rm b}=0.045$, power spectrum normalization $\sigma_{\rm 8}=0.809$, primordial power spectrum index $n_{\rm s}=0.963 $ and Hubble constant $H_0=70.4$ km/s/Mpc. The cosmological parameters are summarized in Table~\ref{tab:cosm_par}. The initial conditions for our simulations were computed using the \cite{Eisenstein:1998p1104} transfer function and the {\scshape grafic++} code developed by Doug Potter (http://sourceforge.net/projects/grafic/) which is based on the original {\scshape grafic} code \citep{Bertschinger:2001p1123}. We first ran a dark matter only simulation with particle mass $m_{\rm cdm}=1.55\times 10^9$~M$_\odot$/h and box size $144$~Mpc/h. The initial level of refinement is $\ell=9$ ($512^3$), but 7 more levels of refinement were carried out during the runs (maximum level $\ell_{\rm max}=16$). 

From this cubic cosmological volume we identified dark matter halos with the AdaptaHOP algorithm \citep{2004MNRAS.352..376A}, using the version implemented and tested by \cite{2009A&A...506..647T}. From the list of identified halos we selected a subsample of 51 halos whose {\it total} masses are $M_{\rm tot}>10^{14}$~M$_\odot$ and whose neighbouring halos do not have masses larger than $M/2$ within a spherical region of five times their virial radius. This choice allowed us to extract high resolution initial conditions to perform re-simulations of these 51 halos. Each halo was re-simulated three times: the first time considering only dark matter, the second and the third time including baryons, but using different physical prescriptions for feedback. Our results are based on these 153 simulations which constitute our cluster catalog. We label the dark matter only simulation as DMO and the hydrodynamical simulations as HYDRO. 

\begin{table*}
{\bfseries Cosmological parameters}
\begin{center}
\begin{tabular}{|l|c|c|c|c|c|c|}
\hline
\hline
 {\itshape Type} & $H_0$ [km s$^{-1}$Mpc$^{-1}$] & $\sigma_{\rm 8}$ & $n_{\rm s} $ & $\Omega_\Lambda$ & $\Omega_{\rm m}$ & $\Omega_{\rm b}$ \\
\hline
\hline
 DMO & 70.4 & 0.809 & 0.963 & 0.728 & 0.272 & - \\
 HYDRO & 70.4 & 0.809 & 0.963 & 0.728 & 0.272 & 0.045 \\
\hline
\hline
\end{tabular}
\end{center}
\caption{Cosmological parameters adopted in our simulations. The DMO label refers to the dark matter only run. The HYDRO label refers to the hydrodynamical runs.}\label{tab:cosm_par}
\end{table*}

In the 51 DMO simulations the dark matter particle mass is $m_{\rm cdm}=1.94\times 10^{8}$~M$_\odot$/h. In the HYDRO runs the dark matter particle mass is $m_{\rm cdm}=1.62\times 10^{8}$~M$_\odot$/h, while the baryon resolution element has a mass of $m_{\rm gas}=3.22\times 10^{7}$~M$_\odot$. In all our simulations we set the maximum refinement level to $\ell=17$ which corresponds to a minimum cell size $\Delta x_{\rm min} = L/2^{\ell_{\rm max}}\simeq 1.07$ kpc/h. The grid was dynamically refined using a quasi-Lagrangian approach: when the dark matter or baryonic mass in a cell reaches 8 times the initial mass resolution, it is split into 8 children cells. The mass and spatial resolution of our simulations is summarized in Table~\ref{tab:mass_par}

In the HYDRO runs, gas dynamics is modeled using a second-order unsplit Godunov scheme \citep{Teyssier:2002p451, Teyssier:2006p413, Fromang:2006p400} based on the HLLC Riemann solver and the
MinMod slope limiter \citep{Toro:1994p1151}. We assume a perfect gas equation of state (EOS) with polytropic index $\gamma=5/3$. All the HYDRO runs include subgrid models for gas cooling which account for H, He and metals and that use the \citealt{sutherland_dopita93} cooling function. We directly follow star formation and supernovae feedback ("delayed cooling" scheme, \citealt{Stinson:2006p1402}) and metal enrichment. In 51 of the HYDRO simulations we also implement AGN feedback, using a modified version of the \cite{Booth:2009p501} model in which supermassive black holes (SMBHs) are modeled as sink particles and AGN feedback is provided in form of thermal energy injected in a sphere surrounding each SMBH. We label these simulations as AGN-ON. The other 51 HYDRO simulations do not include AGN feedback. We label these simulations as AGN-OFF. 

\begin{figure}
    \includegraphics[width=0.49\textwidth]{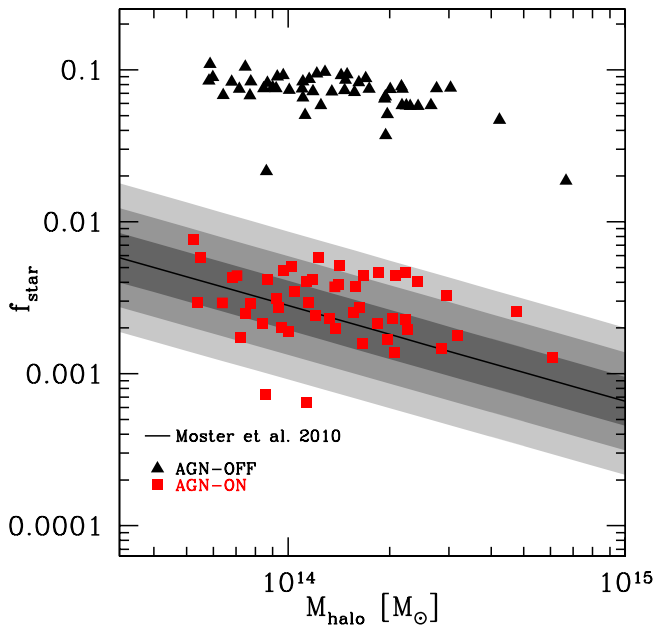}
\caption{ Halo mass vs. stellar fraction associated to the central galaxy of each of our simulated clusters (ratio between galaxy stellar mass and total mass of the halo). Coloured points represent the results from the AGN-ON (red) and AGN-OFF (black) re-simulations. The black solid line represents the average relation from Moster et al. (2010) with its 1, 2 and 3~$\sigma$ scatter (shaded areas).}
  \label{fig:abmatch}
\end{figure}

As we showed in \cite{2012MNRAS.420.2859M}, after implementing the physical effects of AGN feedback we were able to obtain galaxy clusters with correct stellar masses and kinematics. There are a number of unconstrained free parameters in these simulations, for example, the efficiency of the AGN feedback and star formation process. A careful study of the tuning of AGN feedback models implemented in the {\scshape ramses} code has been performed by \cite{2012MNRAS.420.2662D}. In our case, the tuning has been performed re-simulating one of the halos in our catalog (the less massive one) several times while varying the star formation efficiency $\epsilon_*$ and the size of the region where the AGN feedback energy is injected. We found that the model that best reproduces the $M_{\rm BH}-\sigma$ relation and the central galaxy masses has star formation efficiency $\epsilon_*=0.03$ and size of the AGN feedback injection region equal to twice the cell size. For the AGN-OFF simulations we adopt $\epsilon_*=0.01$, which is close to the lower limit of the observed star formation efficiencies. Despite this low value, Agertz et al. (2011) produced realistic "Milky Way" candidates assuming $\epsilon_*=0.01$, however we expect very different results in galaxy clusters because the AGN-OFF simulations are affected by gas overcooling. 

\begin{table}
\begin{center}
{\bfseries Mass and spatial resolution}
\begin{tabular}{|l|c|c|c|}
\hline
\hline
{\itshape Type} & $m_{\rm cdm}$&  $m_{\rm gas}$ & $\Delta x_{\rm min}$ \\
 & $[10^{8}$ M$_\odot$/h] & $[10^{7}$ M$_\odot$/h] & [kpc/h] \\
\hline
\hline
 DMO cube & $ 15.5 $ & n.a. & $2.14$ \\
 DMO clusters & $1.94$ & n.a. & $1.07$ \\
 HYDRO clusters & $1.62$ & $3.22$ & $1.07$ \\
\hline
\hline
\end{tabular}
\end{center}
\caption{Mass resolution for dark matter particles, gas cells and star particles, and spatial resolution (in physical units) for our simulations. }\label{tab:mass_par}
\end{table}

Figure~\ref{fig:abmatch} shows the abundance matching predictions \citep{Moster:2010p5423} for the relation between halo mass and stellar fraction associated to the central galaxy in the clusters, compared to our results. The stellar mass of the central galaxies has been measured as in \cite{2012MNRAS.420.2859M}. Our choice of parameters for the AGN-ON simulations produces results in good agreement with the data. It is worth commenting on the presence of the two outliers in the AGN-ON case: these two objects are highly unrelaxed clusters. In one cluster the BCG is still interacting with a close companion. In the other cluster the BCG is interacting with two companions. It is likely that taking into account the mass in the companions might improve the agreement with the abundance matching prediction. We believe that the selection criterion we adopt to identify the halos used in this study does not bias the following results. As all the unrelaxed clusters, these two outliers were excluded from the analysis concerning the mass function.

\section{Effect of baryonic processes on the halo mass function}
\label{sec:mf_formalism}

From a theoretical perspective, the easiest way to define the mass of a dark matter halo is to use spherical overdensities. We can define the average density within a given radius as:
\begin{equation} \label{mass_def}
 \Delta=\frac{3M(r)}{4\pi r^3}
\end{equation}
where $r$ is the distance from the centre of the halo and $M(r)$ is the spherically averaged enclosed mass profile of the halo. Conventional definitions of the halo mass can be obtained by requiring $\Delta$ to be a multiple of the critical density $\rho_{\rm c}$. For example, by requiring $\Delta=500\rho_{\rm c}$ we define the $M_{500}$ halo mass. In this paper we are interested in studying the effect of baryonic processes related to hydrodynamics and galaxy formation on halo masses, and subsequently on the halo mass function. To do so we adopt a very simple analytical formalism, very similar to the one used by \cite{2013JCAP...04..022B}. 

Cosmological hydrodynamical simulations have been used to show that phenomena related to galaxy formation can strongly influence the internal mass distribution of dark matter halos. Baryon condensation at the centre of halos tends to increase the concentration of the total mass distribution \citep{Gnedin:2004p569}, whereas feedback processes tend to act in the opposite direction by decreasing the concentration through dynamical processes related to the strong winds feedback can drive \citep{2010Natur.463..203G,2010MNRAS.405.2161D,2012ApJ...744L...9M,2012MNRAS.421.3464P,2012MNRAS.422.3081M,2013MNRAS.429.3068T}. It has been shown that the effects of baryons are not limited to changes in the inner density profile, but to global properties such as their mass, e.g. by \cite{2009MNRAS.394L..11S} and \cite{2012MNRAS.423.2279C} and shape, e.g. \cite{2004ApJ...611L..73K}. All of these effects can have important consequences for the halo mass function.

The analytical formalism we adopt is very simple and relies on some approximations that we outline here. We will focus on $M_{500}$ halo masses, but the formalism can be easily adapted to any definition of the halo mass. We have verified that the results we obtain are qualitatively similar in the case in which $M_{200}$ is used. We find that the adoption of $M_{500}$ is more robust because the definition of halo masses is weakly influenced by the presence of structures at the outskirts of the halo. If baryonic effects are important then if we measure $M^{\rm DMO}_{500}$, the mass of a halo in a cosmological dark matter only simulation, the mass measured in a cosmological hydrodynamical simulation will be different, $M^{\rm HYD}_{500}\neq M^{\rm DMO}_{500}$. $M^{\rm HYD}_{500}$ is the sum of the baryonic and non-baryonic mass. We assume the baryon fraction within $R_{500}$ to be a function of halo mass: 
\begin{equation}\label{fb_scaling}
 f_{\rm b, 500}=f_{\rm b, 500}(M^{\rm HYD}_{500})
\end{equation}
which implies that $f_{\rm b, 500}$ is generally different from the cosmological baryon fraction $f_{\rm c}$. In this paper, we adopt a very simple power-law scaling of the baryon fraction with mass:
\begin{equation}\label{pl}
 f_{\rm b, 500}(M^{\rm HYD}_{500})=A\left(\frac{M^{\rm HYD}_{500}}{10^{14} \hbox{M}_{\odot}}\right)^B.
\end{equation}
This choice is convenient since the parameters $A$ and $B$ can be easily measured from simulations as well as from observational data. 

Similar formulae have been used to fit observational data, e.g. by Lin et al. (2003) and Giodini et al. (2009). This has been shown to be appropriate in the range of massive groups and clusters, however significant deviations from a power-law formula have been reported for very massive clusters \citep{2012ApJ...746...95L, 2012ApJ...745L...3L, 2013A&A...555A..66L}. In the mass range we consider in this paper a power-law, like the one in Equation~\ref{fb_scaling}, should correctly capture the scaling of the baryon fractions with halo masses.

The effect of baryonic processes on the halo mass is modeled as a relation between $M^{\rm HYD}_{500}$ and $M^{\rm DMO}_{500}$. To obtain such a relation, \cite{2013JCAP...04..022B} assume that the dark matter halo mass is the same, with or without the presence of baryons. In this paper, we generalize this assumption to take account of the dynamical effect of baryons. Large concentrations of baryonic material at the centre of dark matter halos are expected to generate a contraction of the mass distribution and change the amount of dark matter present in the halos. We parametrize this effect by assuming:
\begin{equation}\label{darkmass}
M_{dark}^{\rm DMO}=\frac{1}{\alpha_c}M_{dark}^{\rm HYD}.
\end{equation}
Here, we have introduced the parameter $\alpha_c$ to account for the fact that the presence of baryons in the halo can induce a steepening of the density distribution within the halo and possibly boost the mass within the virial radius. In the DMO case all the mass is modeled as collisionless dark matter, but we assume that a fraction $f_c$ is associated to baryons. Therefore, in the DMO case we assume a dark matter mass $M_{\rm dark}^{\rm DMO}=(1-f_{\rm c})M_{\rm 500}^{\rm DMO}$, whereas in the HYDRO case we assume $M_{\rm dark}^{\rm HYD}=(1-f_{\rm b,500}(M_{500}^{\rm HYD}))M_{\rm 500}^{\rm HYD}$. This allows us to write Equation~\ref{darkmass} as follows:
\begin{equation}\label{eqn:mass}
 M_{500}^{\rm DMO}(M_{500}^{\rm HYD})=\frac{1}{\alpha_c}\frac{1-f_{\rm b,500}(M_{500}^{\rm HYD})}{1-f_{\rm c}}M_{500}^{\rm HYD}.
\end{equation}
We find that the $\alpha_c$ parameter is quite important for the correct determination of the relation $M_{500}^{\rm DMO}(M_{500}^{\rm HYD})$ and that we cannot simply assume $\alpha_c =1$, i.e. we cannot assume that the dark matter mass is the same in the HYDRO and the DMO case as in \cite{2013JCAP...04..022B}.

In the simplistic case in which the relation between baryon fraction and mass is exact, i.e. when the intrinsic scatter is assumed to be zero, it is straightforward to compute the effect of baryonic processes on the halo mass function. Let us label the mass function affected by baryonic processes as $n(M_{500}^{\rm HYD})$ and the standard mass function measured in dark matter only simulations as $n_0(M_{500}^{\rm HYD})$. Then, if $f_{\rm b, 500}(M^{\rm HYD}_{500})$, we have that
\begin{equation}\label{mf_noscatter}
 n(M_{500}^{\rm HYD})= n_0\left[M_{500}^{\rm DMO}(M_{500}^{\rm HYD})\right]\frac{d M_{500}^{\rm DMO}}{dM_{500}^{\rm HYD}}.
\end{equation}
Equation~\ref{mf_noscatter} can be easily derived by the normalization condition:
$$
\int n_0(M_{500}^{\rm DMO})dM_{500}^{\rm DMO}=\int n(M_{500}^{\rm HYD})dM_{500}^{\rm HYD}.
$$

In the most general case, the relation between baryon fraction and halo mass is not a precise relation due to noise and variance, i.e. it is characterized by an average relation such as the one in Equation~\ref{fb_scaling} and by a scatter about the mean. This fact means that, an $M_{500}^{\rm HYD}$ value is associated to an $M_{500}^{\rm DMO}$ value via a stochastic process. This process is characterized by $P(M_{500}^{\rm HYD}|M_{500}^{\rm DMO})$, the conditional probability density which gives the distribution of the total cluster mass for a given $M_{500}^{\rm DMO}$. In the general case, $n(M_{500}^{\rm HYD})$ has to be computed as 
\begin{equation}\label{mf_nondet}
 n(M_{500}^{\rm HYD})=\int_{0}^{+\infty}n_0(M_{500}^{\rm DMO}) P(M_{500}^{\rm HYD}|M_{500}^{\rm DMO}) dM_{500}^{\rm DMO}.
\end{equation}

In their discussion, \cite{2013JCAP...04..022B} argue that the conditional probability function is very well approximated by a log-normal distribution and we make the same assumption here:
\begin{equation}\label{probability}
 P(M_{500}^{\rm HYD}|M_{500}^{\rm DMO})=\frac{1}{\sqrt{2\pi\sigma^2}}\frac{1}{M_{500}^{\rm HYD}}\exp{\left[-\frac{(\ln M_{500}^{\rm HYD} - \mu)^2}{2\sigma^2}\right]},
\end{equation}
where the parameters $\mu$ and $\sigma$ are defined as
$$
\mu=\ln \langle M_{500}^{\rm HYD} | M_{500}^{\rm DMO}\rangle-\frac{\sigma^2}{2},
$$
$$
\sigma^2=\ln\left(1+ \frac{\sigma^2_{\rm HYD}}{\langle M_{500}^{\rm HYD} | M_{500}^{\rm DMO}\rangle^2} \right),
$$
where $\langle M_{500}^{\rm HYD} | M_{500}^{\rm DMO}\rangle$ is the average expected value of $M_{500}^{\rm HYD}$ given $M_{500}^{\rm DMO}$, and $\sigma^2_{\rm HYD}$ is the scatter about $\langle M_{500}^{\rm HYD} | M_{500}^{\rm DMO}\rangle$. The parameters $\mu$ and $\sigma$ can be easily determined from the results of our simulations. \cite{2013JCAP...04..022B} argue that the log-normal approximation works very well when the log-scatter is a weak function of halo mass. We find that the scatter does not show significant variations in the mass range we are considering, therefore we group all halos in a single bin to compute a unique value for $\sigma^2$. In the AGN-ON case we find $\sigma_{\rm HYD}/\langle M_{500}^{\rm HYD} | M_{500}^{\rm DMO}\rangle=0.046$, whereas in the AGN-OFF case we find $\sigma_{\rm HYD}/\langle M_{500}^{\rm HYD} | M_{500}^{\rm DMO}\rangle=0.065$. 

\section{Baryon fractions and halo masses}

The key elements to compute the effect of baryonic processes on the halo mass function using the formalism described in the previous section are a reliable measure of the $f_{\rm b, 500}(M^{\rm HYD}_{500})$ relation and of its scatter, and a model for the conditional probability function $P(M_{500}^{\rm HYD}|M_{500}^{\rm DMO})$. It is important to note that the halos considered in this paper are less massive than the ones currently used for cosmological analysis, e.g. using all sky ROSAT samples.

We measured halo masses and baryon fractions of all the halos in our cluster catalog and fitted the data using Equation~\ref{fb_scaling} and Equation~\ref{eqn:mass} to obtain an estimate of the average scaling relation $f_{\rm b, 500}(M^{\rm HYD}_{500})$ and of its scatter. All halo masses are estimated using the spherically averaged overdensity definition of Equation~\ref{mass_def}. We then follow the formalism described in the previous section to calculate the expected halo mass functions from the three different simulations. In the following analysis we only consider relaxed clusters in order to avoid biases in the mass measurements that arise from unrelaxed systems. The relaxed clusters are identified through a morphological selection based on mock X-ray emissivity maps. We identify the relaxed clusters by finding those that exhibit a unique emissivity peak within 100 kpc from their center. The selection reduces our cluster sample to 25 relaxed halos. The scatter in the properties of the relaxed sample is significantly reduced, indicating that our relaxation criterion, although minimal, does the job in selecting a uniform, well relaxed population. We show the fitting parameters in Table~\ref{tab:fits}, which have been obtained using a standard $\chi ^2$ algorithm. The reduced chi-squared value $\tilde{\chi} ^2$ for our fits are reported in Table~\ref{tab:fits}. Such values have been obtained by weighting each term of the ${\chi} ^2$ by the square of the inverse of 10\% of the data point value. {The assumed error bars on each mass measurement are supposed to mimic mass calibration errors in real surveys}.

\begin{table*}
{\bfseries Parameters of the mass model}
\begin{center}
\begin{tabular}{|l|c|c|c|c|}
\hline
\hline
 {\itshape Type} & $A$ & $B$ & $\alpha_c$ & $\tilde{\chi} ^2$\\
\hline
\hline
 AGN-ON & $0.1633\pm 0.0012$ & $0.052\pm 0.016$ & $1.039 \pm 0.009$ & 0.56 \\
 AGN-OFF & $0.2038\pm 0.0014$ & $-0.039\pm 0.014$  & $1.156 \pm 0.015$ & 0.37 \\
\hline
\hline
\end{tabular}
\caption{Fits to the simulated data for the baryon fraction scaling and mass model defined in Equation~\ref{pl}. The reduced chi-squared value $\tilde{\chi} ^2$ for the fits is also reported.}\label{tab:fits}
\end{center}
\end{table*}

The scaling of the baryon fraction with cluster mass has been measured in observed clusters by several groups \citep{Lin:2003p1820, Gonzalez:2007p916, Giodini:2009p4283, 2010MNRAS.407..263A, 2012ApJ...746...95L, 2012ApJ...745L...3L}. The debate on which technique provides the best estimate of the baryon fraction scaling with halo mass is still ongoing (weak and strong gravitational lensing, X-ray masses obtained assuming hydrostatic equilibrium, constraints from stellar kinematics, etc.). In this paper, we adopt the fits to the data provided by \cite{Lin:2003p1820} and \cite{Giodini:2009p4283}, which are provided in the form of power laws similar to that in Equation~\ref{pl}. 

\begin{figure}
    \includegraphics[width=0.49\textwidth]{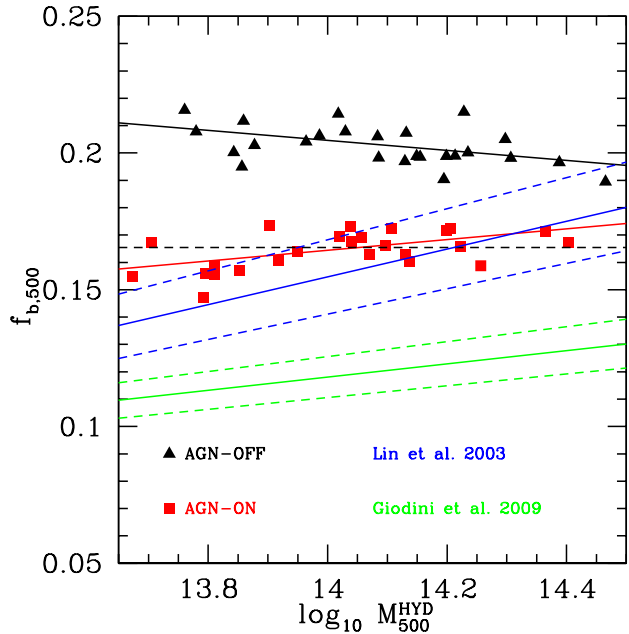}
\caption{Baryon fraction scaling as a function of halo mass at redshift $z=0$. Coloured points represent the results from the AGN-ON (red) and AGN-OFF (black) re-simulations. The red and black solid lines represent power-law fits to the simulations. The blue solid line represents the average relation measured by Lin et al. (2003), whereas the green solid line represents the average relation measured by Giodini et al. (2009). 1-$\sigma$ error bars for the observational results are shown as dashed lines. The black dashed line represents the cosmological baryon fraction.}
  \label{fig:fb}
\end{figure}

Figure~\ref{fig:fb} shows a comparison of the baryon fractions measured within our virialised cluster sample at redshift $z=0$ with the observational results by \cite{Lin:2003p1820} and \cite{Giodini:2009p4283} (re-normalized to fit our choice of cosmological parameters). As expected the AGN-OFF simulations show baryon fractions in excess of the observational results, an effect related to the overcooling of gas in simulations which do not include AGN feedback. 
The tension between observations and simulations is partially relaxed when the AGN-ON case is considered. There is a marginal agreement between the AGN-ON results and the data by \cite{Lin:2003p1820}, whereas our baryon fractions are far in excess of those measured by \cite{Giodini:2009p4283}. In the AGN-ON case, the power law normalization we find is very similar to that found by \cite{Lin:2003p1820}, however their slope is steeper. The normalization found by \cite{Giodini:2009p4283} is lower than in our AGN-ON case, and our slope is shallower. However, since the difference between the observed relations is large, it is not clear if our fits to the simulated data are discrepant with the real universe.

\begin{figure}
    \includegraphics[width=0.49\textwidth]{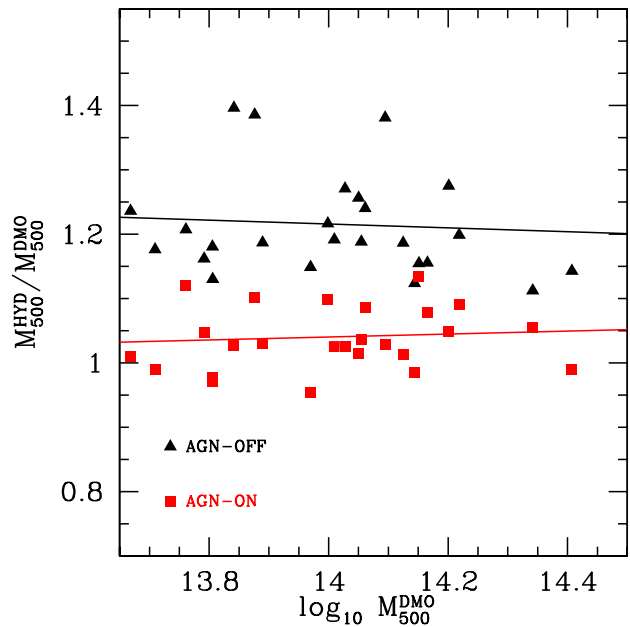}
\caption{Ratio $M_{500}^{\rm HYD}/M_{500}^{\rm DMO}$ as a function of $M_{500}^{\rm DMO}$. Coloured points represent the results from the AGN-ON (red) and AGN-OFF (black) re-simulations. The red and black solid lines represent  fits to the results of our simulations. This plot shows the effect of baryonic processes on the mass of the dark matter halos in our sample.}
  \label{fig:masses}
\end{figure}

\begin{figure}
    \includegraphics[width=0.49\textwidth]{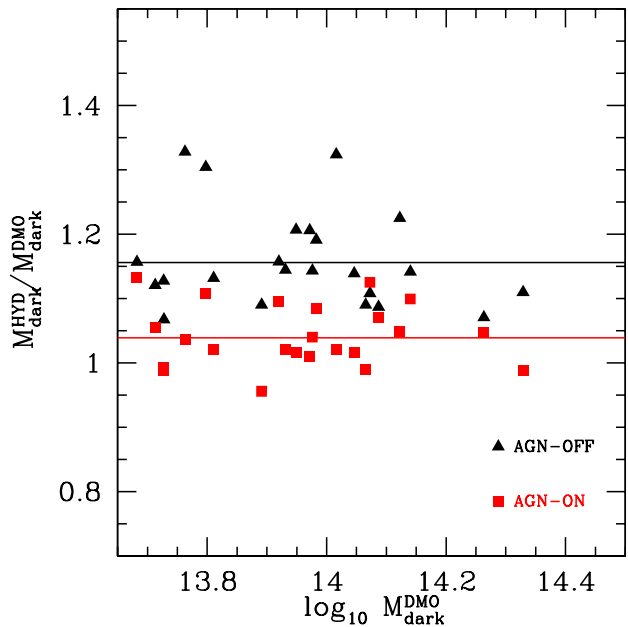}
\caption{$M_{\rm dark}^{\rm DMO}$ vs. ratio $M_{\rm dark}^{\rm HYDRO}/M_{\rm dark}^{\rm DMO}$. Coloured points represent the results from the AGN-ON (red) and AGN-OFF (black) re-simulations. The red and black solid lines represent fits to the results of our analytical model, i.e. the $\alpha_c$ value. }
  \label{fig:darkmasses2}
\end{figure}

The discussion in Section~\ref{sec:mf_formalism} argues that the baryon fraction scaling with mass determines the effects of baryonic processes on the halo mass function. The data shown in Figure~\ref{fig:fb} are relevant to this. In particular, the masses of individual halos are influenced by the nature of the physical processes involved in galaxy formation \citep{2009MNRAS.394L..11S, 2012MNRAS.423.2279C}. This is indeed what we find by analysing our cluster sample, as we show in Figure~\ref{fig:masses} in which we plot the ratio $M_{500}^{\rm HYD}/M_{500}^{\rm DMO}$ as a function of $M_{500}^{\rm DMO}$. This plot demonstrates by how much the masses of the halos in the HYDRO simulations deviate from those measured for the same halos in the DMO runs. We find that in simulations which do not include AGN feedback the halo masses are boosted to values that can be more than 20\% higher than the corresponding DMO value. Previous results obtained with the {\scshape RAMSES} code show that this effect is caused by the contraction of the total mass distribution due to the high concentration of baryons at the centre of the halo \citep{2012MNRAS.422.3081M}. Including AGN feedback acts in the opposite direction, preventing the formation of high concentrations of baryonic mass at the centre of the halo. The net effect is that the halos in the AGN-ON runs have masses that are very similar to the ones measured in the DMO case, although a significant scatter appears to be present in the $M_{500}^{\rm HYD}/M_{500}^{\rm DMO}$ scaling relation. The red and black solid lines in Figure~\ref{fig:masses} show the results of our fit to the data using the model of Equation~\ref{eqn:mass}.

The importance of the contraction parameter $\alpha_c$ can be appreciated in Figure~\ref{fig:darkmasses2} where we plot the ratio $M_{dark}^{\rm HYD}/M_{dark}^{\rm DMO}$ as a function of $M_{dark}^{\rm DMO}$. This plot demonstrates that the slope in the relation observed in Figure~\ref{fig:masses} is set by the variation of the baryon fractions with halo mass. The ratio of dark matter masses $M_{dark}^{\rm HYD}/M_{dark}^{\rm DMO}$ is approximately constant with halo mass in both the AGN-ON and AGN-OFF cases, but it shows a large scatter. The $\alpha_c$ model successfully fits the results observed in our simulations. These results have two important implications: (I) contraction of dark matter induced by baryons has a relevant effect and can be observed only in high resolution simulations; (II) the contraction effect on halo masses can be described by a simple model. 

We have explicitly tested the validity of the adiabatic contraction theory \citep{Gnedin:2004p569} to explain the $\alpha_c$ values we measure in our simulations. We adopt the adiabatic contraction model we already used in \cite{2011MNRAS.414..195T} and \cite{2012MNRAS.422.3081M}, but we assume that all the baryons are concentrated at the centre of the halo (Appendix~\ref{appendix:A}). The adiabatic contraction model predicts $\alpha_c=1.050$ in the AGN-ON case and 
$\alpha_c=1.102$ in the AGN-OFF case. Both values are in good agreement with the results of our simulations, but the AGN-ON halos are slightly under-contracted with respect to the predictions of the adiabatic contraction model, while the AGN-OFF halos are slightly over-contracted.

\section{The halo mass function including baryonic effects}
\label{sec:results}

In the following, we assume that the halo mass function in dark matter only simulations is described by the \cite{2008ApJ...688..709T} formula. The masses used in the \cite{2008ApJ...688..709T} formula are defined in terms of $\rho_m$ the average matter density. To match our mass definitions with those of \cite{2008ApJ...688..709T} we use the relation $500\rho_c=500\rho_m/\Omega_{\rm m}=1838.23\rho_m$. We use the analytical formalism described above and the measurements performed on our re-simulations to calculate the effect of baryons on the halo mass function. 

\begin{figure*}
    \includegraphics[width=0.6\textwidth]{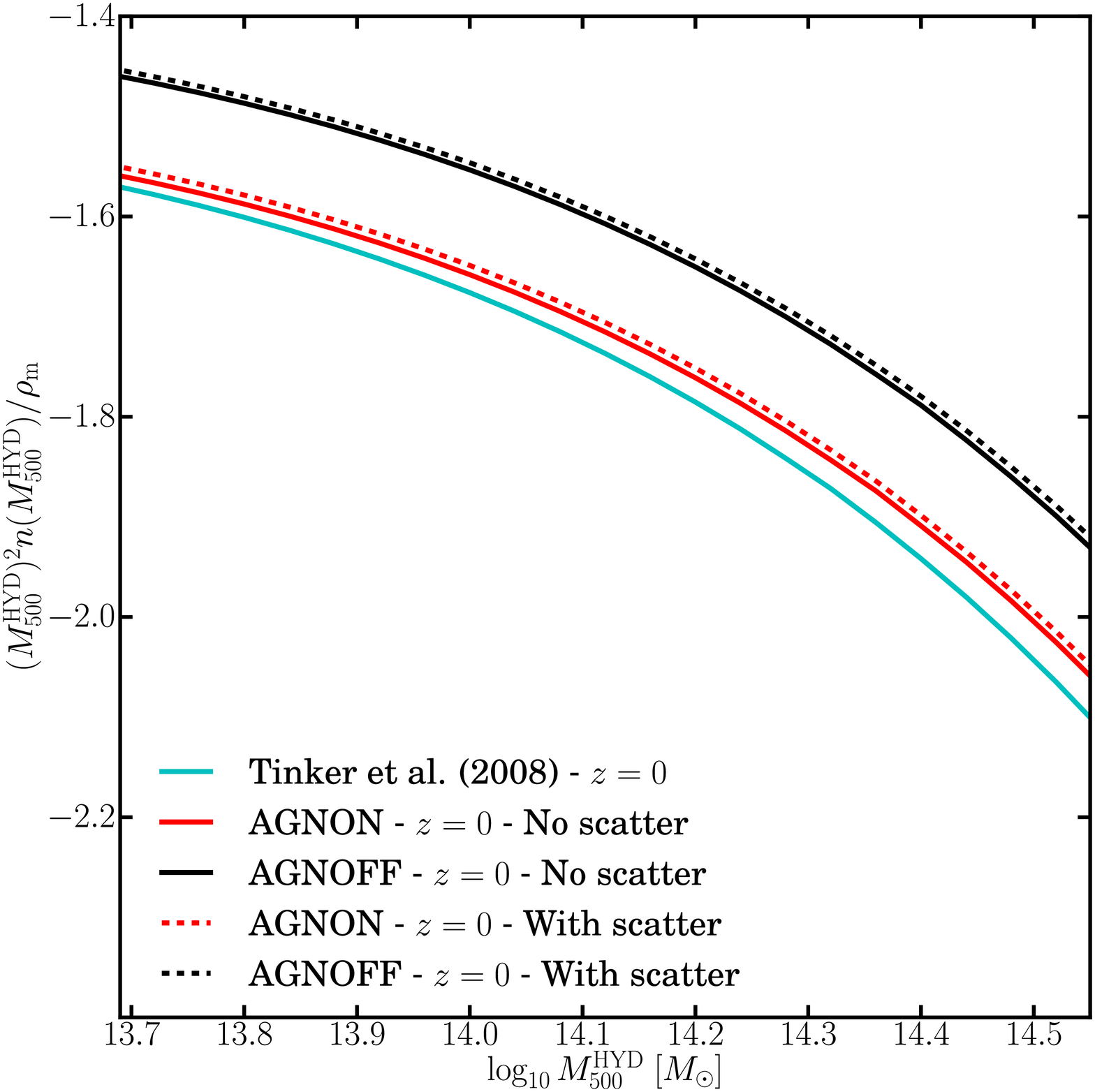}
    \includegraphics[width=0.6\textwidth]{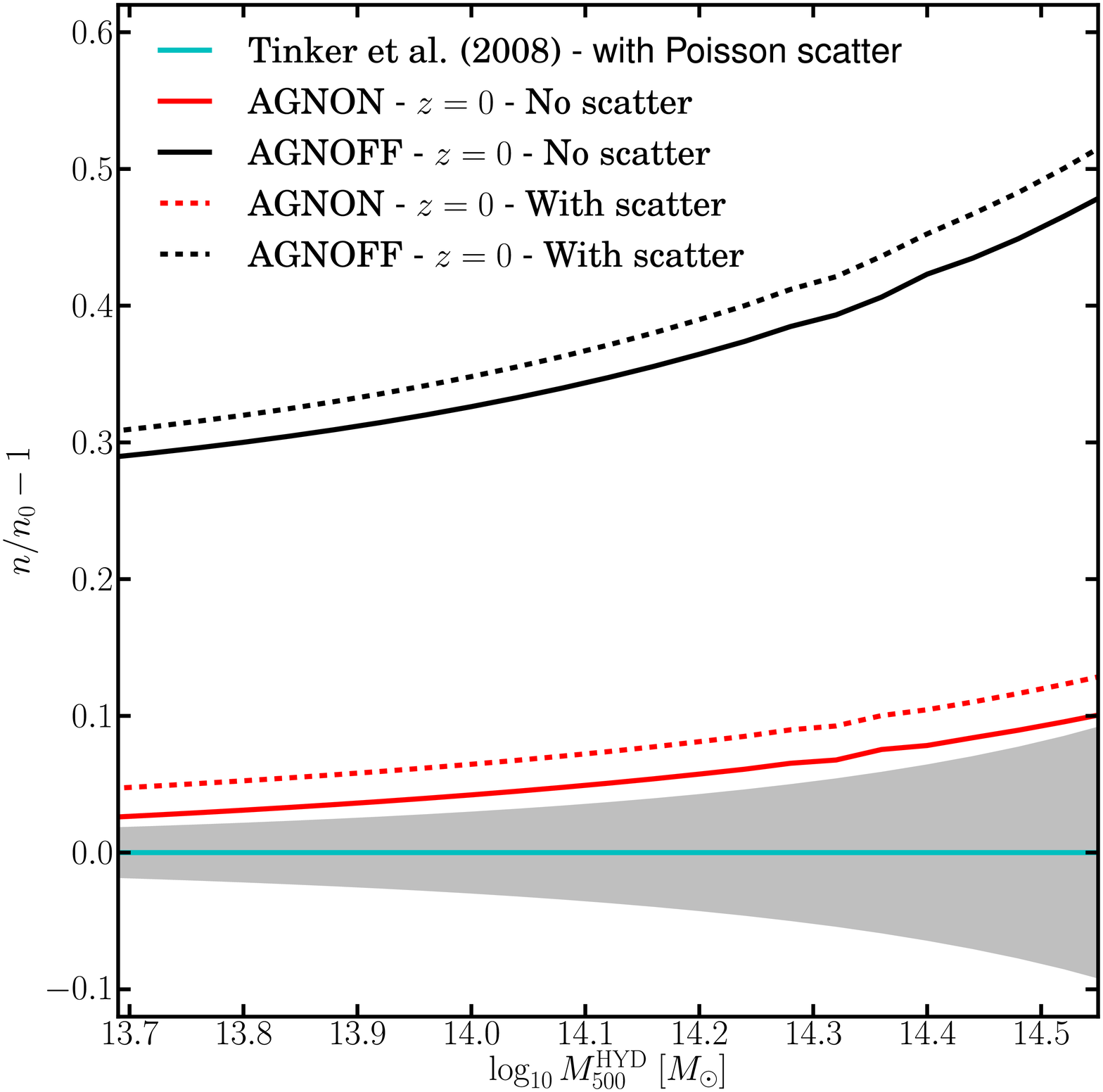}	
\caption{Top panel: The halo mass functions inferred by our model. The cyan solid line is the Tinker et al. (2008) model for the halo mass function in dark matter only simulation. The red and black lines represent the predictions of our model in the AGN-ON and AGN-OFF cases: for the solid lines we assume an exact scaling of the baryon fractions, for the dashed lines we take into account the measured scatter. Bottom panel: Relative deviation of our predictions from the Tinker et al. (2008) model for the halo mass function in dark matter only simulations. The shaded area represents Poisson scatter for a survey of volume 0.5 $h^{-3}$Gpc$^3$. The red and black lines represent the predictions of our model in the AGN-ON and AGN-OFF cases: for the solid lines we assume an exact baryon fraction scaling, for the dashed lines we take into account the measured scatter.}
  \label{fig:mf1n}
\end{figure*}

The comparison between the different results is summarized in Figure~\ref{fig:mf1n}, top panel. We only show the mass range in which the baryon fraction scaling has been measured in our simulations. Figure~\ref{fig:mf1n} clearly shows how the boost in halo masses observed in the AGN-OFF runs influences the halo mass function by increasing the number of halos in the high mass bins. The scatter in the baryon scaling relation has been taken into account by considering the values reported at the end of Section 3 and assuming Equation~\ref{probability} to be valid. The effect of the scatter of the baryon scaling relation boosts the number counts even more. As already discussed, AGN feedback acts in a way that partially suppresses the boost in halo masses observed in AGN-OFF simulations. The net effect is that the prediction for the halo mass function in the AGN-ON case is closer to the dark matter only case. Furthermore, the effect of scatter in the baryon fraction scaling relation is also boosting the number of counts in the AGN-ON case. This boost is simply due to the scatter which populates higher mass bins with lower mass clusters.

There is a small caveat in this determination of the scatter calculated by comparing the same clusters simulated as dark matter only and then with baryons. Some of this scatter will arise from the numerical effects associated with different simulations of a weakly chaotic system. As an extreme example, different simulation codes that start with the same initial conditions will end up with slightly different final masses for the same object.  However, we expect that most of the scatter is physical and arises from the different formation history once baryons are included. For example, cooling and star-formation makes accreting halos denser and able to survive deeper into the potential and taking mass further in. We confirmed that there is a correlation between the mass increase and the stellar mass fraction of the clusters. However, to truly isolate these effects, a large number of clusters with very similar masses should be simulated. 

With the aim of having a more quantitative estimate of how much the mass functions predicted for the HYDRO simulations differ from the dark matter only case, we plot the relative deviation $n/n_{0}-1$ as a function of halo mass in Figure~\ref{fig:mf1n}, bottom panel. We also show the expected Poisson error bars on the halo mass function from a survey of volume 0.5 $h^{-3}$Gpc$^3$ as a grey shaded area in Figure~\ref{fig:mf1n}. This volume is comparable to the volume of currently available galaxy cluster surveys that have been used to obtain cosmological constraints. For instance, the MaxBCG cluster catalogue \citep{2007ApJ...660..239K} covers a volume of $\sim$ 0.17 $h^{-3}$Gpc$^3$ whereas the REFLEX II catalogue \citep{2011MNRAS.413..386B} probes a volume of $\sim$ 2 $h^{-3}$Gpc$^3$. However, such a volume is quite small compared to the one that will be probed by future surveys, e.g. eROSITA, with a covered sky-fraction of 0.658 and redshift limit to be 2.0, will have a survey volume larger than 400 Gpc$^3$. For Euclid the survey volume is expected to be even larger. For this reason the error bars shown in Figure~\ref{fig:mf1n} are much larger than the Poisson noise contribution to the expected uncertainty in the determination of the mass function from future surveys. For larger surveys the Poisson error bars will be much smaller.

The meaning of the error bars in Figure~\ref{fig:mf1n} is worth discussing in more detail. We are making predictions for the halo mass function, which requires halo counts in a given volume. {Cosmological constraints from completed and ongoing surveys are limited by mass calibration and redshift evolution systematics (e.g. \cite{2009ApJ...692.1033V}), therefore Poisson error bars we consider for the smaller volume represent a very idealized case. However, we are not interested in the best possible strategy for cosmological parameters extraction and in the subtleties of the estimation of the error bars, but we are interested in the effect of baryons, therefore our simplified approach should be good enough to study the problem. }

Firstly, we note that the predictions of the AGN-OFF model are well outside the Poisson error bars. In the case in which the scatter in the baryon fraction scaling is explicitly accounted for, the AGN-OFF mass function shows a 30-50\% deviation from the dark matter only prediction of \cite{2008ApJ...688..709T}. If the AGN-OFF model provided the best match to reality, this would imply serious problems for the direct comparison between measured halo mass functions and the results of dark matter only simulations: baryons would need to be explicitly and carefully accounted for when comparing measurements with simulations. However, we already know that the AGN-OFF model is not a good description of observed clusters because it predicts baryon fractions and galaxy masses that are too large. 

The AGN-ON model better matches observations, and the predictions we obtain for the halo mass function are close to the dark matter only case. However, we see that the AGN-ON mass function can still be distinguished from the dark matter only model assuming the Poisson error bars of a very large survey. This result is confirmed also when the scatter in the baryon fraction scaling relation is accounted for. If the results of our simulations are reliable this is a particularly important prediction, since it implies that mass functions directly measured from dark matter only simulations need to be corrected to account for the effect of baryons when comparing to observations. As we discussed above, the main effect of baryons on the halo mass function can be taken into account (e.g. by using Equations~\ref{darkmass} to \ref{mf_nondet}) once the scaling of baryon fractions with mass is known and once a simple analytical model is adopted to account for adiabatic contraction of the dark matter distribution.

\section{Impact on the estimate of cosmological parameters}

In the previous section we showed that including baryonic effects caused the halo mass function to differ from that found in dark matter only simulations. In this section we demonstrate that significant biases can be introduced into the estimation of cosmological parameters when baryon effects are not accounted for. 

We adopt the COSMOMC \citep{2002PhRvD..66j3511L} MCMC to infer $\Omega_{\rm m}$ and $\sigma_{\rm 8}$ whilst assuming a flat universe. In our analysis, we only allowed four parameters to vary: $\Omega_{\rm m}$, $\Omega_{\rm b}$, $h$ and $\sigma_{\rm 8}$. We apply the algorithm to our mass function data using a standard \cite{2008ApJ...688..709T} model to fit the data. We assume Poisson noise to compute the error bars for the resulting halo mass function. The Poisson noise is generated assuming a survey of volume 0.5 $h^{-3}$Gpc$^3$, however in full sky surveys the volumes are larger than this leading to smaller error bars and tighter constraints on the cosmological parameters. To demonstrate that, we also generate Poisson noise for a larger survey of volume 500 $h^{-3}$Gpc$^3$, comparable to eROSITA and Euclid. We stress that Poisson error bars represent an idealized case, however we do not consider this fact important for the significance of our results, as already discussed in the previous section.

\begin{figure*}
    \includegraphics[width=0.6\textwidth]{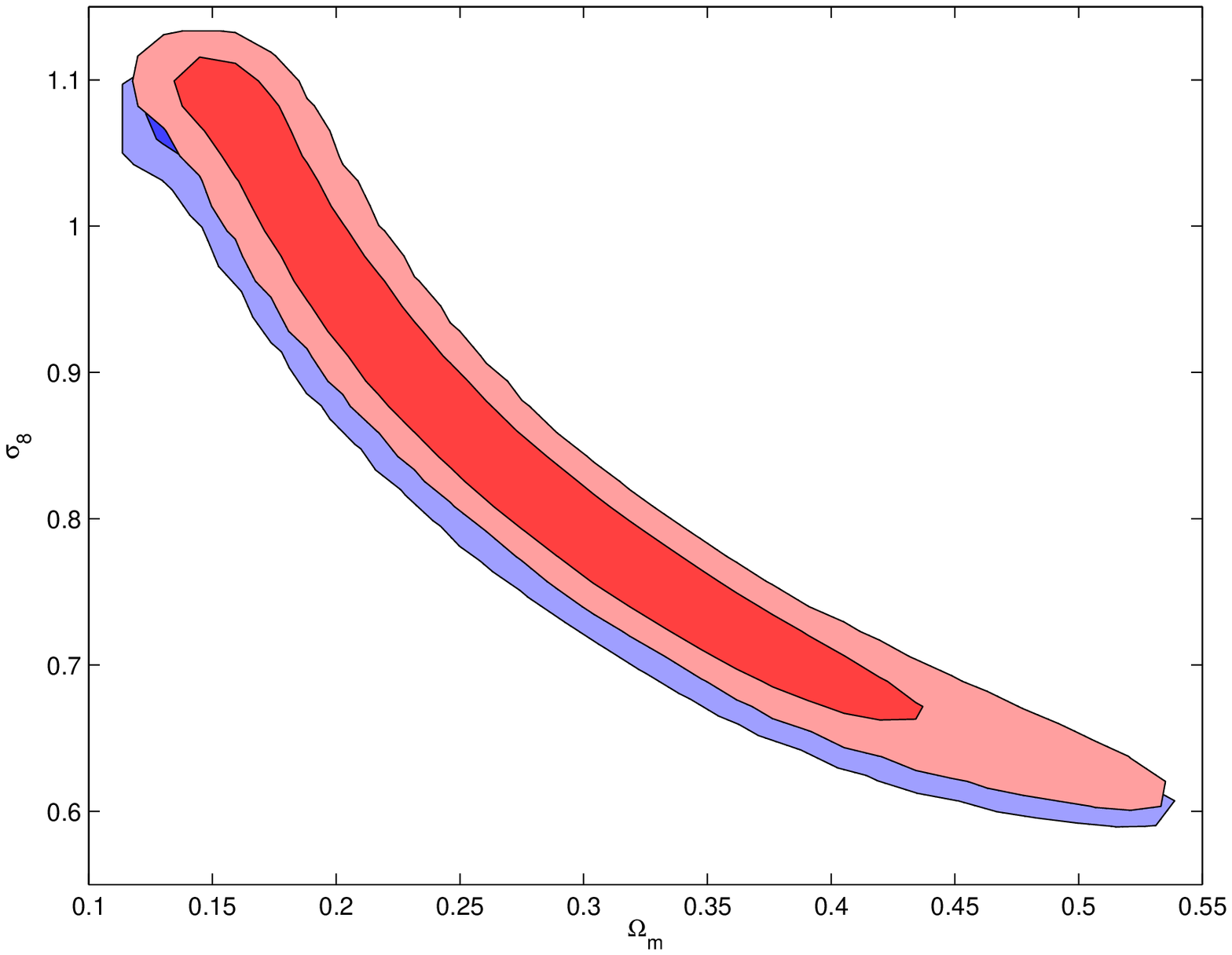}
    \includegraphics[width=0.6\textwidth]{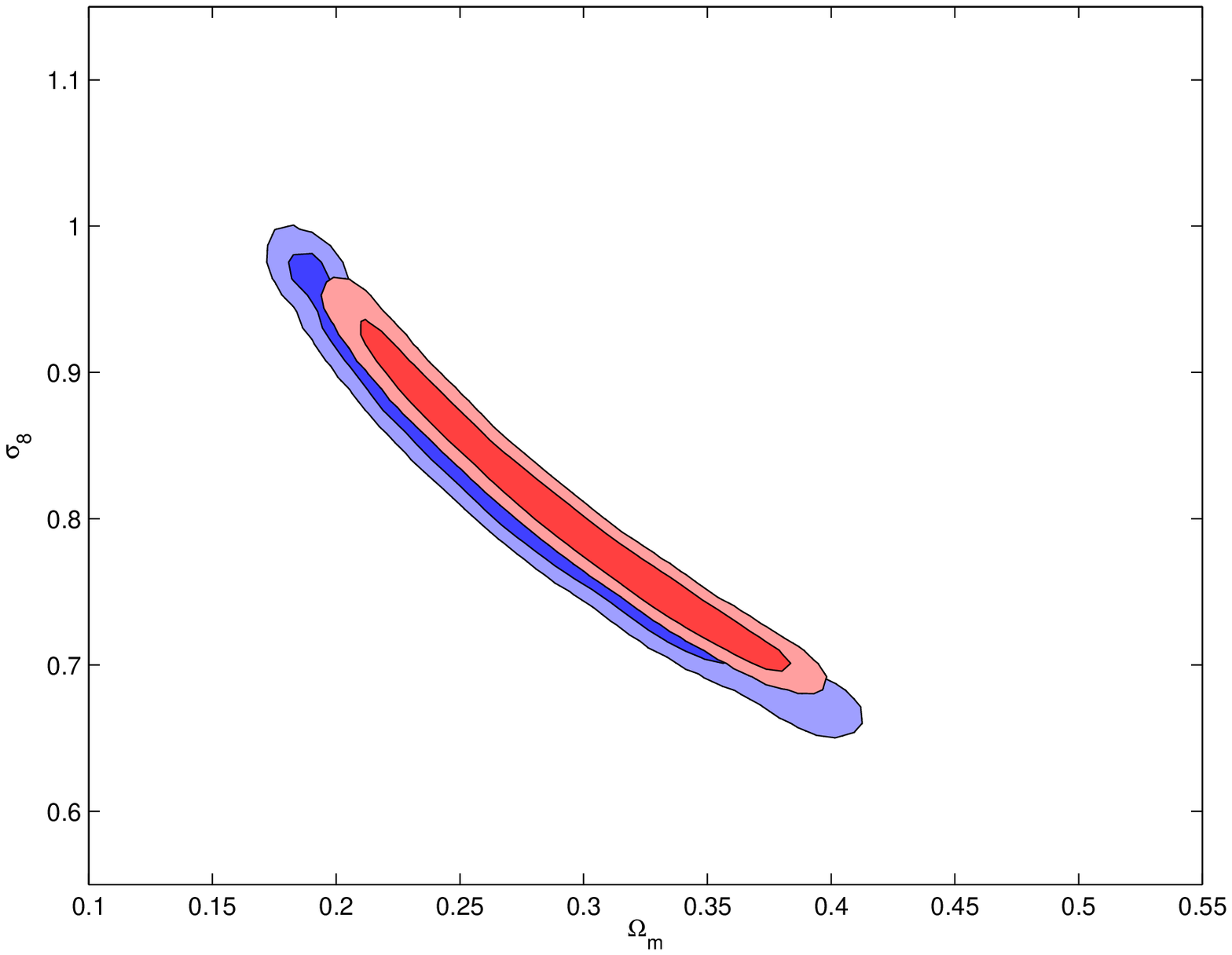}
\caption{Contour plots representing the likelihood as a function of $\Omega_{\rm m}$ and $\sigma_{\rm 8}$. The inner and outer contours are $1\sigma$ and $2\sigma$ away from the peak, respectively. {The blue contours represent the likelihood expected by applying the Tinker et al. (2008) model if the mass function were not influenced by baryonic effects (Table 1). The red contours represent the likelihood in the case in which baryonic effects on the halo mass function are present but the Tinker et al. (2008) model is still adopted.} The top panel represents the case for a survey of volume 0.5 $h^{-3}$Gpc$^3$. The bottom panel represents the case for a survey of volume 500 $h^{-3}$Gpc$^3$.}
  \label{fig:contours}
\end{figure*}

First, we run the MCMC on the \cite{2008ApJ...688..709T} model for the cosmology in Table 1 with the aim of obtaining likelihoods for the cosmological parameters. {If the halo mass function was not influenced by baryonic effects we would get the blue contours in Figure~\ref{fig:contours}, i.e. a likelihood that allows to recover the proper cosmological parameters. Our simulations show that the halo mass function is influenced by baryonic effects, so if we naively applied the \cite{2008ApJ...688..709T} model to the halo mass function modified by baryonic effects we would obtain a likelihood that is biased with respect to the blue contours, because the \cite{2008ApJ...688..709T} model neglects baryonic effects. This biased likelihood is represented by the red contours in Figure~\ref{fig:contours}, obtained by running the MCMC on our AGN-ON model.} Figure~\ref{fig:contours} clearly demonstrates the existence of such a bias. The top panel is for a survey of volume 0.5 $h^{-3}$Gpc$^3$, the bottom panel is for a survey of volume 500 $h^{-3}$Gpc$^3$. In the smaller volume, we find a bias in the $\Omega_{\rm m}$ and $\sigma_{\rm 8}$ parameters, at the $0.4\%$ and $1.7\%$ level respectively. In the larger, volume we find a bias in the $\Omega_{\rm m}$ and $\sigma_{\rm 8}$ parameters, at the $7\%$ and $1.5\%$ level respectively. In the bottom panel, the constraints are tighter than in the top panel (due to a larger volume), so the shift is noticeable. Because the case with baryons has a slightly higher mass function, Poisson bars are smaller and hence the constraints are also tighter. In the top panel the volume is smaller and hence the Poisson bars are larger, so the difference is not easily noticeable. Given the high degeneracy between the $\Omega_{\rm m}$ and $\sigma_{\rm 8}$ parameters the overall bias is higher, up to few \%. This bias will be more significant if tighter constraints are obtained from full sky surveys, as shown in the bottom panel of Figure~\ref{fig:contours}, and if the constraints coming from the halo mass function are combined with those obtained from other probes (CMB, SN-Ia, BAO). If sub-percent accuracy cosmology is the aim of the next generation observational campaign, baryonic effects clearly need to be taken into account. 

\section{Summary and Conclusions}
\label{sec:summary}

We identified 51 isolated dark matter halos in a cosmological dark matter only simulation and re-simulated them at higher resolution using different prescriptions for galaxy formation. The {\itshape total} halo masses are $10^{14}$~M$_\odot< M_{\rm tot}<10^{15}$~M$_\odot$, therefore they host massive groups and clusters of galaxies. In the AGN-OFF model we implement standard galaxy formation recipes, including supernovae feedback. In the AGN-ON model we also include AGN feedback as in \cite{Booth:2009p501}. We measured the masses and baryon fractions of the halos in the catalog and we used our formalism to compute how the halo mass function changes under the effect of baryons. We adopted the \cite{2008ApJ...688..709T} mass function as our fiducial analytical model for the halo mass function measured in dark matter only simulations. We then used a simple analytical formalism to compute the effect of baryons on the dark matter halo mass function at redshift $z=0$ and we applied it to obtain predictions from cosmological hydrodynamical simulations.

Our findings can be summarized in a few points:
\begin{itemize}

 \item The AGN-ON model provides better fits to the observations. The masses of the central galaxies in the clusters are in agreement with the values expected from abundance matching \citep{Moster:2010p5423, 2013MNRAS.428.3121M}. The scaling of baryon fractions with halo mass is in relatively good agreement with the results published by \cite{Lin:2003p1820}, but in excess with respect to the values measured by \cite{Giodini:2009p4283}. The AGN-OFF model produces central galaxies that are too massive and baryon fractions that are too high compared to the observational results. Therefore, the AGN-ON model produces more realistic results with respect to the case without AGN feedback.

 \item In the AGN-OFF case halo masses can be $>20$\% larger than in the dark matter only runs. This is an effect of the condensation of baryonic mass at the centre of the dark matter halos, which also boosts the concentration of the total mass distribution \citep{2012MNRAS.422.3081M}. AGN feedback acts in the opposite direction, preventing large condensations of baryonic mass at the centre of dark matter halos. The halo masses in the AGN-ON runs are very close to those measured in the dark matter only case.

 \item Halo counts in the sampled mass range are boosted by $\sim 30-50$\% in the AGN-OFF case. In the more realistic AGN-ON case we found that the halo mass function for massive groups and clusters is still not consistent with the halo mass function measured in dark matter only simulations. The halo mass function in the AGN-ON case is boosted by 5-10 \%. This fact implies that mass functions measured in dark matter only simulations may not be accurate enough to be directly compared to observational results. 

\item Neglecting baryonic effects on the halo mass function leads to biases at the \% level in the estimate of cosmological parameters like $\Omega_{\rm m}$ and $\sigma_{\rm 8}$. If we assume a survey volume of 0.5 $h^{-3}$Gpc$^3$, comparable to present the one of current surveys, we find a relevant bias in the $\Omega_{\rm m}$ and $\sigma_{\rm 8}$ parameters, at the $0.4\%$ and $1.7\%$ level, respectively.

\item Modeling of the baryonic effects on the halo mass function requires to explicitly account for the contraction of the dark matter distribution under the effect of baryons. We propose an extension of the model used by \cite{2013JCAP...04..022B}, which uses the baryon fraction scaling with halo mass, parameterises the response of dark matter to baryonic condensations at the halo centres and predicts the effect of baryons on halo masses.

\end{itemize}

There are a few caveats that should be discussed. First of all, we measured properties only for a limited number of halos. The bias between the pure dark matter halo mass function and that found with a treatment of hydrodynamics, star-formation, supernovae and AGN feedback, may be present at the few percent level. To improve these estimates we need more statistics and a more detailed exploration of the parameter space. The best approach would be to consider a full cosmological box instead of several re-simulations. However, simulations of full cosmological boxes with the resolution we achieve and including AGN feedback are still computationally challenging. This might serve as a partial justification for the adoption of the semi-analytical approach we follow in this paper. Furthermore, the problem of studying the baryonic effects on the halo mass functions has also been recently studied by \cite{2014arXiv1402.4461V} using SPH simulations; despite the difference of their results with ours, their conclusions is still the same: baryonic effects on the halo mass function are important. The predictions for the halo mass function including baryonic effects need to be carefully studied with larger and improved simulations, since these effects will then be relevant. 

Furthermore, our analytical model to correct the mass function for baryonic effects has two limitations: firstly, the baryon fraction scaling and its scatter needs to be known; secondly, a simplified model which accounts for adiabatic contraction needs to be adopted. The former limitation could be overcome using accurate observations of the baryon fraction scaling. The latter limitation represents an intrinsic theoretical issue - the adiabatic contraction theory is only an approximation to account for the response of the mass distribution to large baryon condensations. Our results seem to confirm that such an approximation is reasonable, since the features of the halos in our numerical simulations are well captured by a simple adiabatic contraction model. We also point out that alternative techniques that take into account the effect of baryons on the halo mass function are being developed by other authors (see e.g. \cite{2013arXiv1309.4094C}).


Finally, we want to stress that our results have significant implications for the correct comparison of halo counts in present and next generation SZ (e.g. Planck, South Pole Telescope, Atacama Cosmology Telescope), X-ray (e.g. Chandra, XMM-Newton, Suzaku, eROSITA) and multi-wavelength surveys (e.g. Dark Energy Survey, Euclid): the effects of baryons influence the halo mass function at the cluster scale at a level $\sim10$\% or less, which is larger than the level of accuracy that next generation surveys should achieve. At the galaxy cluster scale, the proper modeling of baryon physics is still likely to be an issue for what concerns the comparison between the mass function measured from observations and the one measured in simulations. For current cluster surveys baryonic effects are within the noise for current survey volumes. Recent constraints placed on cosmological parameters using cluster surveys (e.g. ROSAT \citep{2011MNRAS.413..386B, 2013MNRAS.432..973R}, SDSS \citep{2012arXiv1207.3794Z, 2013MNRAS.434..684M}, ACT \citep{2013JCAP...07..008H}, SPT \citep{2013ApJ...763..127R, 2013ApJ...763..147B}, PLANCK \citep{2013arXiv1303.5080P})  are not expected to be strongly influenced by baryonic effects, but forthcoming and planned large surveys will be biased at the \% level by these processes.

\section*{Acknowledgments}

We thank Darren Reed for very useful discussion on the topic. We also thank the anonymous referee whose comments greatly improved the quality of this paper. All the simulations were performed on the Cray XE6 cluster Monte Rosa at CSCS, Lugano, Switzerland.


\bibliography{papers}

\begin{thebibliography}{}

\bibitem[\protect\citeauthoryear{{Andreon}}{{Andreon}}{2010}]{2010MNRAS.407..263A}
{Andreon} S.,  2010, \mnras, 407, 263

\bibitem[\protect\citeauthoryear{{Aubert}, {Pichon} \& {Colombi}}{{Aubert}
  et~al.}{2004}]{2004MNRAS.352..376A}
{Aubert} D.,  {Pichon} C.,    {Colombi} S.,  2004, \mnras, 352, 376

\bibitem[\protect\citeauthoryear{{Audit}, {Teyssier} \& {Alimi}}{{Audit}
  et~al.}{1997}]{1997A&A...325..439A}
{Audit} E.,  {Teyssier} R.,    {Alimi} J.-M.,  1997, \aap, 325, 439

\bibitem[\protect\citeauthoryear{{Balaguera-Antolinez} \&
  {Porciani}}{{Balaguera-Antolinez} \& {Porciani}}{2013}]{2013JCAP...04..022B}
{Balaguera-Antolinez} A.,  {Porciani} C.,  2013, Jcap, 4, 22

\bibitem[\protect\citeauthoryear{{Balaguera-Antol{\'{\i}}nez}, {S{\'a}nchez},
  {B{\"o}hringer}, {Collins}, {Guzzo} \& {Phleps}}{{Balaguera-Antol{\'{\i}}nez}
  et~al.}{2011}]{2011MNRAS.413..386B}
{Balaguera-Antol{\'{\i}}nez} A.,  {S{\'a}nchez} A.~G.,  {B{\"o}hringer} H.,
  {Collins} C.,  {Guzzo} L.,    {Phleps} S.,  2011, \mnras, 413, 386

\bibitem[\protect\citeauthoryear{{Benson}, {de Haan}, {Dudley}, {Reichardt},
  {Aird}, {Andersson}, {Armstrong}, {Ashby}, {Bautz}, {Bayliss}, {Bazin},
  {Bleem}, {Brodwin}, {Carlstrom} \& {Chang}}{{Benson}
  et~al.}{2013}]{2013ApJ...763..147B}
{Benson} B.~A.,  {de Haan} T.,  {Dudley} J.~P.,  {Reichardt} C.~L.,  {Aird}
  K.~A.,  {Andersson} K.,  {Armstrong} R.,  {Ashby} M.~L.~N.,  {Bautz} M.,
  {Bayliss} M.,  {Bazin} G.,  {Bleem} L.~E.,  {Brodwin} M.,  {Carlstrom} J.~E.,
     {Chang} C.~L. e.~a.,  2013, \apj, 763, 147

\bibitem[\protect\citeauthoryear{Bertschinger}{Bertschinger}{2001}]{Bertschinger:2001p1123}
Bertschinger E.,  2001, The Astrophysical Journal Supplement Series, 137, 1

\bibitem[\protect\citeauthoryear{Booth \& Schaye}{Booth \&
  Schaye}{2009}]{Booth:2009p501}
Booth C.~M.,  Schaye J.,  2009, Monthly Notices of the Royal Astronomical
  Society, 398, 53

\bibitem[\protect\citeauthoryear{{Crocce}, {Fosalba}, {Castander} \&
  {Gazta{\~n}aga}}{{Crocce} et~al.}{2010}]{2010MNRAS.403.1353C}
{Crocce} M.,  {Fosalba} P.,  {Castander} F.~J.,    {Gazta{\~n}aga} E.,  2010,
  \mnras, 403, 1353

\bibitem[\protect\citeauthoryear{{Cui}, {Borgani}, {Dolag}, {Murante} \&
  {Tornatore}}{{Cui} et~al.}{2012}]{2012MNRAS.423.2279C}
{Cui} W.,  {Borgani} S.,  {Dolag} K.,  {Murante} G.,    {Tornatore} L.,  2012,
  \mnras, 423, 2279

\bibitem[\protect\citeauthoryear{{Cusworth}, {Kay}, {Battye} \&
  {Thomas}}{{Cusworth} et~al.}{2013}]{2013arXiv1309.4094C}
{Cusworth} S.~J.,  {Kay} S.~T.,  {Battye} R.~A.,    {Thomas} P.~A.,  2013,
  ArXiv e-prints arXiv:1309.4094

\bibitem[\protect\citeauthoryear{{Dubois}, {Devriendt}, {Slyz} \&
  {Teyssier}}{{Dubois} et~al.}{2012}]{2012MNRAS.420.2662D}
{Dubois} Y.,  {Devriendt} J.,  {Slyz} A.,    {Teyssier} R.,  2012, \mnras, 420,
  2662

\bibitem[\protect\citeauthoryear{{Duffy}, {Schaye}, {Kay}, {Dalla Vecchia},
  {Battye} \& {Booth}}{{Duffy} et~al.}{2010}]{2010MNRAS.405.2161D}
{Duffy} A.~R.,  {Schaye} J.,  {Kay} S.~T.,  {Dalla Vecchia} C.,  {Battye}
  R.~A.,    {Booth} C.~M.,  2010, \mnras, 405, 2161

\bibitem[\protect\citeauthoryear{{Efstathiou}, {Frenk}, {White} \&
  {Davis}}{{Efstathiou} et~al.}{1988}]{1988MNRAS.235..715E}
{Efstathiou} G.,  {Frenk} C.~S.,  {White} S.~D.~M.,    {Davis} M.,  1988,
  \mnras, 235, 715

\bibitem[\protect\citeauthoryear{Eisenstein \& Hu}{Eisenstein \&
  Hu}{1998}]{Eisenstein:1998p1104}
Eisenstein D.~J.,  Hu W.,  1998, Astrophysical Journal v.496, 496, 605

\bibitem[\protect\citeauthoryear{Fromang, Hennebelle \& Teyssier}{Fromang
  et~al.}{2006}]{Fromang:2006p400}
Fromang S.,  Hennebelle P.,    Teyssier R.,  2006, Astronomy and Astrophysics,
  457, 371

\bibitem[\protect\citeauthoryear{Giodini, Pierini, Finoguenov, Pratt,
  Boehringer, Leauthaud, Guzzo, Aussel, Bolzonella \& the
  COSMOS~Collaboration}{Giodini et~al.}{2009}]{Giodini:2009p4283}
Giodini S.,  Pierini D.,  Finoguenov A.,  Pratt G.~W.,  Boehringer H.,
  Leauthaud A.,  Guzzo L.,  Aussel H.,  Bolzonella M.,    the
  COSMOS~Collaboration 2009, The Astrophysical Journal, 703, 982

\bibitem[\protect\citeauthoryear{Gnedin, Kravtsov, Klypin \& Nagai}{Gnedin
  et~al.}{2004}]{Gnedin:2004p569}
Gnedin O.~Y.,  Kravtsov A.~V.,  Klypin A.~A.,    Nagai D.,  2004, The
  Astrophysical Journal, 616, 16

\bibitem[\protect\citeauthoryear{Gonzalez, Zaritsky \& Zabludoff}{Gonzalez
  et~al.}{2007}]{Gonzalez:2007p916}
Gonzalez A.~H.,  Zaritsky D.,    Zabludoff A.~I.,  2007, The Astrophysical
  Journal, 666, 147

\bibitem[\protect\citeauthoryear{{Governato}, {Brook}, {Mayer}, {Brooks},
  {Rhee}, {Wadsley}, {Jonsson}, {Willman}, {Stinson}, {Quinn} \&
  {Madau}}{{Governato} et~al.}{2010}]{2010Natur.463..203G}
{Governato} F.,  {Brook} C.,  {Mayer} L.,  {Brooks} A.,  {Rhee} G.,  {Wadsley}
  J.,  {Jonsson} P.,  {Willman} B.,  {Stinson} G.,  {Quinn} T.,    {Madau} P.,
  2010, \nat, 463, 203

\bibitem[\protect\citeauthoryear{{Guillet}, {Teyssier} \& {Colombi}}{{Guillet}
  et~al.}{2010}]{2010MNRAS.405..525G}
{Guillet} T.,  {Teyssier} R.,    {Colombi} S.,  2010, \mnras, 405, 525

\bibitem[\protect\citeauthoryear{{Hasselfield}, {Hilton}, {Marriage},
  {Addison}, {Barrientos}, {Battaglia}, {Battistelli}, {Bond} \&
  {Crichton}}{{Hasselfield} et~al.}{2013}]{2013JCAP...07..008H}
{Hasselfield} M.,  {Hilton} M.,  {Marriage} T.~A.,  {Addison} G.~E.,
  {Barrientos} L.~F.,  {Battaglia} N.,  {Battistelli} E.~S.,  {Bond} J.~R.,
  {Crichton} D. e.~a.,  2013, Jcap, 7, 8

\bibitem[\protect\citeauthoryear{{Jenkins}, {Frenk}, {White}, {Colberg},
  {Cole}, {Evrard}, {Couchman} \& {Yoshida}}{{Jenkins}
  et~al.}{2001}]{2001MNRAS.321..372J}
{Jenkins} A.,  {Frenk} C.~S.,  {White} S.~D.~M.,  {Colberg} J.~M.,  {Cole} S.,
  {Evrard} A.~E.,  {Couchman} H.~M.~P.,    {Yoshida} N.,  2001, \mnras, 321,
  372

\bibitem[\protect\citeauthoryear{{Jing}, {Zhang}, {Lin}, {Gao} \&
  {Springel}}{{Jing} et~al.}{2006}]{2006ApJ...640L.119J}
{Jing} Y.~P.,  {Zhang} P.,  {Lin} W.~P.,  {Gao} L.,    {Springel} V.,  2006,
  \apjl, 640, L119

\bibitem[\protect\citeauthoryear{{Kazantzidis}, {Kravtsov}, {Zentner},
  {Allgood}, {Nagai} \& {Moore}}{{Kazantzidis}
  et~al.}{2004}]{2004ApJ...611L..73K}
{Kazantzidis} S.,  {Kravtsov} A.~V.,  {Zentner} A.~R.,  {Allgood} B.,  {Nagai}
  D.,    {Moore} B.,  2004, \apjl, 611, L73

\bibitem[\protect\citeauthoryear{{Koester}, {McKay}, {Annis}, {Wechsler},
  {Evrard}, {Bleem}, {Becker}, {Johnston}, {Sheldon}, {Nichol}, {Miller} \&
  {Scranton}}{{Koester} et~al.}{2007}]{2007ApJ...660..239K}
{Koester} B.~P.,  {McKay} T.~A.,  {Annis} J.,  {Wechsler} R.~H.,  {Evrard} A.,
  {Bleem} L.,  {Becker} M.,  {Johnston} D.,  {Sheldon} E.,  {Nichol} R.,
  {Miller} C.,    {Scranton} R. e.~a.,  2007, \apj, 660, 239

\bibitem[\protect\citeauthoryear{{Lagan{\'a}}, {Martinet}, {Durret}, {Lima
  Neto}, {Maughan} \& {Zhang}}{{Lagan{\'a}} et~al.}{2013}]{2013A&A...555A..66L}
{Lagan{\'a}} T.~F.,  {Martinet} N.,  {Durret} F.,  {Lima Neto} G.~B.,
  {Maughan} B.,    {Zhang} Y.-Y.,  2013, \aap, 555, A66

\bibitem[\protect\citeauthoryear{{Leauthaud}, {George}, {Behroozi}, {Bundy},
  {Tinker}, {Wechsler}, {Conroy}, {Finoguenov} \& {Tanaka}}{{Leauthaud}
  et~al.}{2012}]{2012ApJ...746...95L}
{Leauthaud} A.,  {George} M.~R.,  {Behroozi} P.~S.,  {Bundy} K.,  {Tinker} J.,
  {Wechsler} R.~H.,  {Conroy} C.,  {Finoguenov} A.,    {Tanaka} M.,  2012,
  \apj, 746, 95

\bibitem[\protect\citeauthoryear{{Lewis} \& {Bridle}}{{Lewis} \&
  {Bridle}}{2002}]{2002PhRvD..66j3511L}
{Lewis} A.,  {Bridle} S.,  2002, \prd, 66, 103511

\bibitem[\protect\citeauthoryear{Lin, Mohr \& Stanford}{Lin
  et~al.}{2003}]{Lin:2003p1820}
Lin Y.-T.,  Mohr J.~J.,    Stanford S.~A.,  2003, The Astrophysical Journal,
  591, 749

\bibitem[\protect\citeauthoryear{{Lin}, {Stanford}, {Eisenhardt}, {Vikhlinin},
  {Maughan} \& {Kravtsov}}{{Lin} et~al.}{2012}]{2012ApJ...745L...3L}
{Lin} Y.-T.,  {Stanford} S.~A.,  {Eisenhardt} P.~R.~M.,  {Vikhlinin} A.,
  {Maughan} B.~J.,    {Kravtsov} A.,  2012, \apjl, 745, L3

\bibitem[\protect\citeauthoryear{{Macci{\`o}}, {Stinson}, {Brook}, {Wadsley},
  {Couchman}, {Shen}, {Gibson} \& {Quinn}}{{Macci{\`o}}
  et~al.}{2012}]{2012ApJ...744L...9M}
{Macci{\`o}} A.~V.,  {Stinson} G.,  {Brook} C.~B.,  {Wadsley} J.,  {Couchman}
  H.~M.~P.,  {Shen} S.,  {Gibson} B.~K.,    {Quinn} T.,  2012, \apjl, 744, L9

\bibitem[\protect\citeauthoryear{{Mana}, {Giannantonio}, {Weller}, {Hoyle},
  {H{\"u}tsi} \& {Sartoris}}{{Mana} et~al.}{2013}]{2013MNRAS.434..684M}
{Mana} A.,  {Giannantonio} T.,  {Weller} J.,  {Hoyle} B.,  {H{\"u}tsi} G.,
  {Sartoris} B.,  2013, \mnras, 434, 684

\bibitem[\protect\citeauthoryear{{Martizzi}, {Teyssier} \& {Moore}}{{Martizzi}
  et~al.}{2012}]{2012MNRAS.420.2859M}
{Martizzi} D.,  {Teyssier} R.,    {Moore} B.,  2012, \mnras, 420, 2859

\bibitem[\protect\citeauthoryear{{Martizzi}, {Teyssier}, {Moore} \&
  {Wentz}}{{Martizzi} et~al.}{2012}]{2012MNRAS.422.3081M}
{Martizzi} D.,  {Teyssier} R.,  {Moore} B.,    {Wentz} T.,  2012, \mnras, 422,
  3081

\bibitem[\protect\citeauthoryear{{Moster}, {Naab} \& {White}}{{Moster}
  et~al.}{2013}]{2013MNRAS.428.3121M}
{Moster} B.~P.,  {Naab} T.,    {White} S.~D.~M.,  2013, \mnras, 428, 3121

\bibitem[\protect\citeauthoryear{Moster, Somerville, Maulbetsch, van~den Bosch,
  Macci{\`o}, Naab \& Oser}{Moster et~al.}{2010}]{Moster:2010p5423}
Moster B.~P.,  Somerville R.~S.,  Maulbetsch C.,  van~den Bosch F.~C.,
  Macci{\`o} A.~V.,  Naab T.,    Oser L.,  2010, The Astrophysical Journal,
  710, 903

\bibitem[\protect\citeauthoryear{{Planck Collaboration}, {Ade}, {Aghanim},
  {Armitage-Caplan}, {Arnaud}, {Ashdown}, {Atrio-Barandela}, {Aumont},
  {Baccigalupi} \& {Banday}}{{Planck Collaboration}
  et~al.}{2013}]{2013arXiv1303.5080P}
{Planck Collaboration} {Ade} P.~A.~R.,  {Aghanim} N.,  {Armitage-Caplan} C.,
  {Arnaud} M.,  {Ashdown} M.,  {Atrio-Barandela} F.,  {Aumont} J.,
  {Baccigalupi} C.,    {Banday} A.~J. e.~a.,  2013, ArXiv e-prints
  arXiv:1303.5080

\bibitem[\protect\citeauthoryear{{Pontzen} \& {Governato}}{{Pontzen} \&
  {Governato}}{2012}]{2012MNRAS.421.3464P}
{Pontzen} A.,  {Governato} F.,  2012, \mnras, 421, 3464

\bibitem[\protect\citeauthoryear{{Press} \& {Schechter}}{{Press} \&
  {Schechter}}{1974}]{1974ApJ...187..425P}
{Press} W.~H.,  {Schechter} P.,  1974, \apj, 187, 425

\bibitem[\protect\citeauthoryear{{Rapetti}, {Blake}, {Allen}, {Mantz},
  {Parkinson} \& {Beutler}}{{Rapetti} et~al.}{2013}]{2013MNRAS.432..973R}
{Rapetti} D.,  {Blake} C.,  {Allen} S.~W.,  {Mantz} A.,  {Parkinson} D.,
  {Beutler} F.,  2013, \mnras, 432, 973

\bibitem[\protect\citeauthoryear{{Reddick}, {Tinker}, {Wechsler} \&
  {Lu}}{{Reddick} et~al.}{2013}]{2013arXiv1306.4686R}
{Reddick} R.,  {Tinker} J.,  {Wechsler} R.,    {Lu} Y.,  2013, ArXiv e-prints
  arXiv:1306.4686

\bibitem[\protect\citeauthoryear{{Reed}, {Gardner}, {Quinn}, {Stadel},
  {Fardal}, {Lake} \& {Governato}}{{Reed} et~al.}{2003}]{2003MNRAS.346..565R}
{Reed} D.,  {Gardner} J.,  {Quinn} T.,  {Stadel} J.,  {Fardal} M.,  {Lake} G.,
    {Governato} F.,  2003, \mnras, 346, 565

\bibitem[\protect\citeauthoryear{{Reed}, {Bower}, {Frenk}, {Jenkins} \&
  {Theuns}}{{Reed} et~al.}{2007}]{2007MNRAS.374....2R}
{Reed} D.~S.,  {Bower} R.,  {Frenk} C.~S.,  {Jenkins} A.,    {Theuns} T.,
  2007, \mnras, 374, 2

\bibitem[\protect\citeauthoryear{{Reed}, {Smith}, {Potter}, {Schneider},
  {Stadel} \& {Moore}}{{Reed} et~al.}{2013}]{2013MNRAS.431.1866R}
{Reed} D.~S.,  {Smith} R.~E.,  {Potter} D.,  {Schneider} A.,  {Stadel} J.,
  {Moore} B.,  2013, \mnras, 431, 1866

\bibitem[\protect\citeauthoryear{{Reichardt}, {Stalder}, {Bleem}, {Montroy},
  {Aird}, {Andersson}, {Armstrong}, {Ashby}, {Bautz}, {Bayliss}, {Bazin},
  {Benson}, {Brodwin} \& {Carlstrom}}{{Reichardt}
  et~al.}{2013}]{2013ApJ...763..127R}
{Reichardt} C.~L.,  {Stalder} B.,  {Bleem} L.~E.,  {Montroy} T.~E.,  {Aird}
  K.~A.,  {Andersson} K.,  {Armstrong} R.,  {Ashby} M.~L.~N.,  {Bautz} M.,
  {Bayliss} M.,  {Bazin} G.,  {Benson} B.~A.,  {Brodwin} M.,    {Carlstrom}
  J.~E. e.~a.,  2013, \apj, 763, 127

\bibitem[\protect\citeauthoryear{{Rudd}, {Zentner} \& {Kravtsov}}{{Rudd}
  et~al.}{2008}]{2008ApJ...672...19R}
{Rudd} D.~H.,  {Zentner} A.~R.,    {Kravtsov} A.~V.,  2008, \apj, 672, 19

\bibitem[\protect\citeauthoryear{{Semboloni}, {Hoekstra}, {Schaye}, {van
  Daalen} \& {McCarthy}}{{Semboloni} et~al.}{2011}]{2011MNRAS.417.2020S}
{Semboloni} E.,  {Hoekstra} H.,  {Schaye} J.,  {van Daalen} M.~P.,
  {McCarthy} I.~G.,  2011, \mnras, 417, 2020

\bibitem[\protect\citeauthoryear{{Sheth}, {Mo} \& {Tormen}}{{Sheth}
  et~al.}{2001}]{2001MNRAS.323....1S}
{Sheth} R.~K.,  {Mo} H.~J.,    {Tormen} G.,  2001, \mnras, 323, 1

\bibitem[\protect\citeauthoryear{{Sheth} \& {Tormen}}{{Sheth} \&
  {Tormen}}{1999}]{1999MNRAS.308..119S}
{Sheth} R.~K.,  {Tormen} G.,  1999, \mnras, 308, 119

\bibitem[\protect\citeauthoryear{{Springel}}{{Springel}}{2005}]{2005MNRAS.364.1105S}
{Springel} V.,  2005, \mnras, 364, 1105

\bibitem[\protect\citeauthoryear{{Stanek}, {Rudd} \& {Evrard}}{{Stanek}
  et~al.}{2009}]{2009MNRAS.394L..11S}
{Stanek} R.,  {Rudd} D.,    {Evrard} A.~E.,  2009, \mnras, 394, L11

\bibitem[\protect\citeauthoryear{Stinson, Seth, Katz, Wadsley, Governato \&
  Quinn}{Stinson et~al.}{2006}]{Stinson:2006p1402}
Stinson G.,  Seth A.,  Katz N.,  Wadsley J.,  Governato F.,    Quinn T.,  2006,
  Monthly Notices of the Royal Astronomical Society, 373, 1074

\bibitem[\protect\citeauthoryear{Sutherland \& Dopita}{Sutherland \&
  Dopita}{1993}]{sutherland_dopita93}
Sutherland R.~S.,  Dopita M.~A.,  1993, \apjs, 88, 253

\bibitem[\protect\citeauthoryear{Teyssier}{Teyssier}{2002}]{Teyssier:2002p451}
Teyssier R.,  2002, Astronomy and Astrophysics, 385, 337

\bibitem[\protect\citeauthoryear{Teyssier, Fromang \& Dormy}{Teyssier
  et~al.}{2006}]{Teyssier:2006p413}
Teyssier R.,  Fromang S.,    Dormy E.,  2006, Journal of Computational Physics,
  218, 44

\bibitem[\protect\citeauthoryear{{Teyssier}, {Moore}, {Martizzi}, {Dubois} \&
  {Mayer}}{{Teyssier} et~al.}{2011}]{2011MNRAS.414..195T}
{Teyssier} R.,  {Moore} B.,  {Martizzi} D.,  {Dubois} Y.,    {Mayer} L.,  2011,
  \mnras, 414, 195

\bibitem[\protect\citeauthoryear{{Teyssier}, {Pontzen}, {Dubois} \&
  {Read}}{{Teyssier} et~al.}{2013}]{2013MNRAS.429.3068T}
{Teyssier} R.,  {Pontzen} A.,  {Dubois} Y.,    {Read} J.~I.,  2013, \mnras,
  429, 3068

\bibitem[\protect\citeauthoryear{{Tinker}, {Kravtsov}, {Klypin}, {Abazajian},
  {Warren}, {Yepes}, {Gottl{\"o}ber} \& {Holz}}{{Tinker}
  et~al.}{2008}]{2008ApJ...688..709T}
{Tinker} J.,  {Kravtsov} A.~V.,  {Klypin} A.,  {Abazajian} K.,  {Warren} M.,
  {Yepes} G.,  {Gottl{\"o}ber} S.,    {Holz} D.~E.,  2008, \apj, 688, 709

\bibitem[\protect\citeauthoryear{Toro, Spruce \& Speares}{Toro
  et~al.}{1994}]{Toro:1994p1151}
Toro E.~F.,  Spruce M.,    Speares W.,  1994, Shock Waves, 4, 25

\bibitem[\protect\citeauthoryear{{Tweed}, {Devriendt}, {Blaizot}, {Colombi} \&
  {Slyz}}{{Tweed} et~al.}{2009}]{2009A&A...506..647T}
{Tweed} D.,  {Devriendt} J.,  {Blaizot} J.,  {Colombi} S.,    {Slyz} A.,  2009,
  \aap, 506, 647

\bibitem[\protect\citeauthoryear{{van Daalen}, {Schaye}, {Booth} \& {Dalla
  Vecchia}}{{van Daalen} et~al.}{2011}]{2011MNRAS.415.3649V}
{van Daalen} M.~P.,  {Schaye} J.,  {Booth} C.~M.,    {Dalla Vecchia} C.,  2011,
  \mnras, 415, 3649

\bibitem[\protect\citeauthoryear{{Velliscig}, {van Daalen}, {Schaye},
  {McCarthy}, {Cacciato}, {Le Brun} \& {Dalla Vecchia}}{{Velliscig}
  et~al.}{2014}]{2014arXiv1402.4461V}
{Velliscig} M.,  {van Daalen} M.~P.,  {Schaye} J.,  {McCarthy} I.~G.,
  {Cacciato} M.,  {Le Brun} A.~M.~C.,    {Dalla Vecchia} C.,  2014, ArXiv
  e-prints 1402.4461

\bibitem[\protect\citeauthoryear{{Vikhlinin}, {Burenin}, {Ebeling}, {Forman},
  {Hornstrup}, {Jones}, {Kravtsov}, {Murray}, {Nagai}, {Quintana} \&
  {Voevodkin}}{{Vikhlinin} et~al.}{2009}]{2009ApJ...692.1033V}
{Vikhlinin} A.,  {Burenin} R.~A.,  {Ebeling} H.,  {Forman} W.~R.,  {Hornstrup}
  A.,  {Jones} C.,  {Kravtsov} A.~V.,  {Murray} S.~S.,  {Nagai} D.,  {Quintana}
  H.,    {Voevodkin} A.,  2009, \apj, 692, 1033

\bibitem[\protect\citeauthoryear{{Warren}, {Abazajian}, {Holz} \&
  {Teodoro}}{{Warren} et~al.}{2006}]{2006ApJ...646..881W}
{Warren} M.~S.,  {Abazajian} K.,  {Holz} D.~E.,    {Teodoro} L.,  2006, \apj,
  646, 881

\bibitem[\protect\citeauthoryear{{White}}{{White}}{2004}]{2004APh....22..211W}
{White} M.,  2004, Astroparticle Physics, 22, 211

\bibitem[\protect\citeauthoryear{{Zhan} \& {Knox}}{{Zhan} \&
  {Knox}}{2004}]{2004ApJ...616L..75Z}
{Zhan} H.,  {Knox} L.,  2004, \apjl, 616, L75

\bibitem[\protect\citeauthoryear{{Zu}, {Weinberg}, {Rozo}, {Sheldon}, {Tinker}
  \& {Becker}}{{Zu} et~al.}{2012}]{2012arXiv1207.3794Z}
{Zu} Y.,  {Weinberg} D.~H.,  {Rozo} E.,  {Sheldon} E.~S.,  {Tinker} J.~L.,
  {Becker} M.~R.,  2012, ArXiv e-prints arXiv:1207.3794

\end{thebibliography}


\appendix
\section{Adiabatic Contraction Model}\label{appendix:A}

The simplified model we adopt is based on that used in \cite{2011MNRAS.414..195T}. If one defines the initial radius of each dark matter shell as $r_{\rm i}$, then an adiabatic contraction model is able to predict its value after contraction $r_{\rm f}$. In our case we adopt the transformation
\begin{equation}\label{eq:AC}
 \frac{r_{\rm f}}{r_{\rm i}}=1+\alpha\left(\frac{M_{\rm i}}{M_{\rm f}}-1\right)
\end{equation}
where $M_{\rm i}$ and $M_{\rm f}$ are the cumulative mass distributions before and after contraction, respectively. The final cumulative mass distribution can be computed as
\begin{equation}\label{eq:AC1}
 M_{\rm f}=M_{\rm dark}(r_{\rm f})+M_{\rm b}(r_{\rm f})=f_{\rm dark}M_{\rm i}(r_{\rm i})+M_{\rm b}(r_{\rm f}).
\end{equation}
where $M_{\rm i}(r_{\rm i})$ is the initial mass distribution in the DMO case, $M_{\rm b}(r_{\rm f})$ is the baryonic mass distribution and $M_{\rm dark}(r_{\rm f})$ is the adiabatically contracted dark matter distribution. The dark mass fraction is computed as $f_{\rm dark}=1-f_{\rm b}$, where $f_{\rm b}$ is the baryon fraction. Our aim is to recover the contracted dark matter profile $M_{\rm dark}(r_{\rm f})$ given $M_{\rm i}(r_{\rm i})$ and $M_{\rm b}(r_{\rm f})$. We measure $M_{\rm i}(r_{\rm i})$ from our simulations. As a simplifying assumptions for our simple calculations, we assume that all the baryons are located at the halo centre:
\begin{equation}
M_{\rm b}(r_{\rm f})=M_{\rm 0b}\delta(r_{\rm f}),
\end{equation}
where $\delta(r_{\rm f})$ is Dirac's delta and $M_{\rm 0b}$ is the total baryonic mass. 

We obtain the $r_{\rm f}$ value associated to each $r_{\rm i}$ solving numerically Equation \ref{eq:AC}, and naturally obtain the adiabatically contracted dark matter profile $M_{\rm dm}(r_{\rm f})$ using Equation \ref{eq:AC1}. Once the contracted profile is known, it is easy to measure the new virial mass and radius. 

\label{lastpage}
\end{document}